\documentclass[
 reprint,
superscriptaddress,
 amsmath,amssymb,
aip,
jcp,
citeautoscript
]{revtex4-2}

\usepackage{graphicx}
\usepackage{dcolumn}
\usepackage{bm}
\usepackage{xcolor}
\usepackage[version=4]{mhchem}
\usepackage{xcolor}
\usepackage{braket}
\usepackage{balance}
\usepackage{float}
\usepackage{lipsum}

\begin{document}


%
%

\title{Spectroscopic signatures of biexcitons: A case study in Ruddlesden-Popper lead-halides}

\author{Katherine A. Koch}
\affiliation{Department of Physics and Center for Functional Materials, Wake Forest University, 2090 Eure Drive, Winston-Salem, NC~27109, United~States}

\author{Esteban~Rojas-Gatjens}
\affiliation{School of Chemistry and Biochemistry, Georgia Institute of Technology, 901 Atlantic Drive, Atlanta, GA~30332, United~States}


\author{Mart\'in~G\'omez-Dominguez}%
\affiliation{School of Materials Science and Engineering, Georgia Institute of Technology, 771 Ferst Dr NW, Atlanta, GA~30332, United~States}%

\author{Juan-Pablo~Correa-Baena}
\affiliation{School of Chemistry and Biochemistry, Georgia Institute of Technology, 901 Atlantic Drive, Atlanta, GA~30332, United~States}%
\affiliation{School of Materials Science and Engineering, Georgia Institute of Technology, 771 Ferst Dr NW, Atlanta, GA~30332, United~States}

\author{Carlos~Silva-Acu\~na}
\affiliation{School of Chemistry and Biochemistry, Georgia Institute of Technology, 901 Atlantic Drive, Atlanta, GA~30332, United~States}
\affiliation{School of Materials Science and Engineering, Georgia Institute of Technology, 771 Ferst Dr NW, Atlanta, GA~30332, United~States}
\affiliation{Institut Courtois \& D\'epartement de Physique, Universit\'e de Montr\'eal, 1375 Avenue Th\'er\`ese-Lavoie-Roux, Montr\'eal H2V~0B3, Qu\'ebec, Canada}

\author{Ajay~Ram~Srimath~Kandada}
\affiliation{Department of Physics and Center for Functional Materials, Wake Forest University, 2090 Eure Drive, Winston-Salem, NC~27109, United~States}
\email{srimatar@wfu.edu}

\date{\today}

\begin{abstract}
Exciton-exciton interactions are fundamental to the light-emitting properties of semiconductors, influencing applications from lasers to quantum light sources. In this study, we investigate the spectroscopic signatures and binding energy of biexcitons in a metal halide two-dimensional Ruddlesden-Popper structure, which is known for hosting distinct excitonic resonances with unique lattice coupling. Using three spectroscopic techniques—photoluminescence (PL) and two variations of two-dimensional electronic spectroscopy (2DES)—we map coherent one-quantum and two-quantum correlations to gain deeper insight into the biexciton characteristics. While PL spectroscopy is hindered by spectral broadening and reabsorption, 2DES provides a more accurate characterization, revealing multiple biexciton states and uncovering a mixed biexciton species arising from exciton cross-coupling. These findings highlight the importance of advanced spectroscopic approaches in accurately determining biexciton binding energies and offer new perspectives on many-body interactions in exciton-polarons within layered perovskites.
\end{abstract}

\maketitle

\section{Introduction \label{sec:intro}}

A key factor influencing the light-emitting properties of a semiconducting material is the strength of exciton-exciton interactions. In highly excited materials, these interactions cause a shift in the emission frequency of a biexciton (a bound pair of excitons) compared to a single exciton. This biexciton shift is sometimes harnessed in applications such as lasers~\cite{booker2018vertical, masumoto1993biexciton, grim2014continuous, wang2019ultralow, kondo1998biexciton, he2022multicolor}, where it facilitates population inversion in the biexciton-to-exciton transition, or in the realization of Bose-Einstein condensation of polaritons~\cite{polimeno2020observation, chase1979evidence}. However, in light-emitting applications, biexcitons can reduce efficiency at target wavelengths or compromise the quantum purity of emitted photons. 

The ability to tune the biexciton binding energy is highly desirable, and the first step in this process is to accurately determine its value using various spectroscopic techniques. The most straightforward approach is photoluminescence (PL) spectroscopy, where the energy difference between the exciton and biexciton emission peaks provides a direct estimate of the biexciton binding energy~\cite{birkedal1996binding, cho2024size, hayakawa2014binding}. Alternatively, transient absorption (TA) spectroscopy can be used, in which the energy of the photo-induced absorption feature corresponding to the exciton-to-biexciton transition serves the same purpose~\cite{styers2008exciton, zhang2016understanding, shukla2020effect, sewall2008state, yumoto2018hot}. However, these widely used methods face significant challenges due to strong spectral overlap between multiple broad features, resulting in spectral congestion. Additionally, the unambiguous assignment of spectral features to biexcitons remains problematic, as other species—such as defect states, self-trapped excitons, or various excited-state species—may exhibit similar energy signatures, further complicating the estimation. Many-body interactions in material systems that host multiple excitons give rise to multiple biexciton-like species, making the photophysics even more complex. These simple spectroscopic techniques are often insufficient to resolve such intricacies.

We employ two-dimensional electronic spectroscopy (2DES) as an alternative approach to overcome these limitations and to elucidate the spectroscopic signatures of biexcitons. As a case study, we investigate two-dimensional Ruddlesden-Popper metal halides—prototypical excitonic systems known for their complex many-body interactions. Notably, the optical excitations in these materials can be described as exciton-polarons, where the electron-hole pairs are dressed by lattice phonons. Our previous findings demonstrated that this unique lattice coupling gives rise to multiple coexisting excitonic resonances in the optical spectra separated by approximately 35–40 meV~\cite{thouin2019enhanced, thouin2019phonon, srimath2020exciton}. These exciton states exhibit distinct many-body interactions and biexciton binding characteristics~\cite{thouin2018stable}. 

While our previous studies have primarily focused on phenethylammonium lead iodide (\ce{(PEA)2PbI4}), here we investigate its fluorinated counterpart, in which the organic cation \ce{PEA} is replaced by \ce{F-PEA}, featuring a fluorine substitution at the para-position. In a recent study~\cite{koch2025fine}, we observed that this substitution induces subtle yet distinct changes in the lattice parameters, particularly in the octahedral distortions of the metal-halide network. While these structural modifications are not pronounced enough to significantly alter the electronic structure or exciton binding energy, they have a remarkable impact on exciton-lattice interactions. Notably, we find that these changes manifest in measurable variations in the exciton spectral lineshapes observed in optical absorption, primarily due to modifications in the Huang-Rhys parameters. Furthermore, as reported in Ref.~\citenum{koch2025fine}, this substitution also affects many-body scattering rates. A key motivation of this study is to explore whether multi-exciton dynamics are similarly influenced by organic-inorganic interactions within the lattice, and our findings indeed suggest that such interactions play a role.




\section{Results and Analysis\label{sec:results}}

\begin{figure*}
    \centering
    \includegraphics[width=17cm]{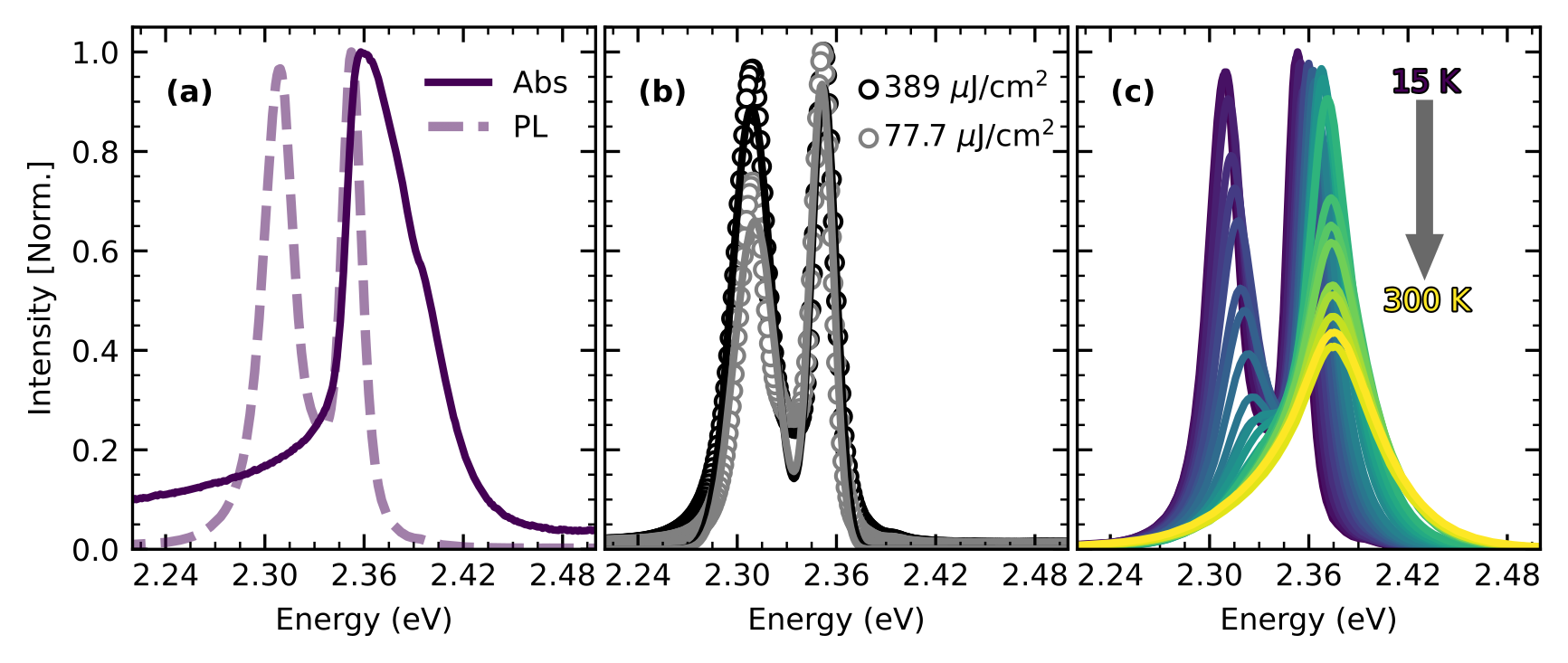}
    \caption{(a) Absorption (solid) and photoluminescence (dashed) spectra, of \ce{(F-PEA)2PbI4} measured at 15 K. (b) Measured photoluminescence spectra (dots) of \ce{(F-PEA)2PbI4} taken at 15 K, fit to a double Gaussian function (lines) to determine peak positions and calculate the biexciton binding energy, see Fig.~\ref{fig:biX_bind}(c). (c) Temperature dependent photoluminesence, measured with a pump fluence of 389 $\mu$J/cm$^2$.} 
    \label{fig:abs_PL}
\end{figure*}

The linear absorption spectrum of the polycrystalline thin film of \ce{(F-PEA)2PbI4}, measured at 15 K, is shown in Fig.\ref{fig:abs_PL}(a). It exhibits a characteristic excitonic peak, accompanied by a broad shoulder at higher energies, likely due to overlapping contributions from a higher-lying excitonic resonance, as discussed in Ref.\citenum{srimath2020exciton, koch2025fine}. Also shown in Fig.\ref{fig:abs_PL} is the photoluminescence (PL) spectrum obtained from the same sample via photo-excitation of carriers in the continuum. The PL spectrum features two well-resolved resonances: the primary peak, exhibiting a relatively small Stokes shift from the absorption peak, corresponds to the radiative recombination of excitons. Additionally, a red-shifted peak appears at around 2.3 eV, whose intensity increases with higher excitation fluence, as shown in Fig.\ref{fig:abs_PL}(b). Given its fluence dependence, previous studies have attributed this peak to the relaxation of a biexciton state, which preferentially forms at higher carrier densities before decaying into the exciton state~\cite{fujisawa2004excitons, kaiser2021free, goto2006localization, makino2005induced}. 
Consequently, this simple PL measurement provides a means to estimate biexciton binding energies, which have been reported to be approximately 40–50\,meV in similar material systems.

To accurately determine the biexciton binding energy from the PL spectra, we fit the PL spectrum with a double Gaussian function (solid lines in Fig.~\ref{fig:abs_PL}(b)), extracting the peak positions of the two resonances. The energy difference between these peaks provides an estimate of the binding energy, which, in this case, is 44\,meV. This value aligns well with reported values for two-dimensional perovskite derivatives and follows the empirical relationship between exciton and biexciton binding energies~\cite{thouin2018stable, ishihara1992dielectric, kondo1998biexciton, kato2003extremely}. As shown in Fig.~\ref{fig:abs_PL}(a), the Stokes shift is negligible and within the linewidths of both the absorption and PL peaks. This suggests that the higher-energy components of the PL spectrum may be reabsorbed by the sample, leading to an underestimation of the biexciton binding energy. Additionally, we observe that the biexciton peak exhibits a broader linewidth compared to the exciton peak. While this may indicate distinct thermal dephasing processes affecting exciton and biexcitonic resonances, the temperature-dependent PL spectra suggest a more intricate scenario.

Fig.~\ref{fig:abs_PL}(c) shows that as the temperature increases, the intensity of the biexciton resonance decreases, and its peak shifts closer to the exciton resonance, implying a reduction in biexciton binding energy. However, the broad linewidths, particularly at elevated temperatures, prevent us from drawing definitive conclusions. These observations highlight how reabsorption effects and broad spectral features obscure the biexciton photophysics, demonstrating that linear spectroscopy alone is insufficient for a complete analysis.

To overcome the limitations of linear spectroscopy in studying biexciton photophysics, we employ two-dimensional electronic spectroscopy (2DES), which offers distinct advantages by resolving both spectral and temporal dynamics that linear methods cannot capture. Unlike absorption and photoluminescence (PL) spectroscopy, which provide ensemble-averaged spectral information, 2DES differentiates between homogeneous and inhomogeneous broadening, offering a clearer view of excited-state energies. It directly maps exciton-biexciton couplings, which are often obscured in linear spectra due to overlapping features and reabsorption effects. Furthermore, 2DES tracks coherence dynamics and many-body interactions over time, providing critical insights into dephasing mechanisms. 
\begin{figure}
    \centering
    \includegraphics[width=8.5cm]{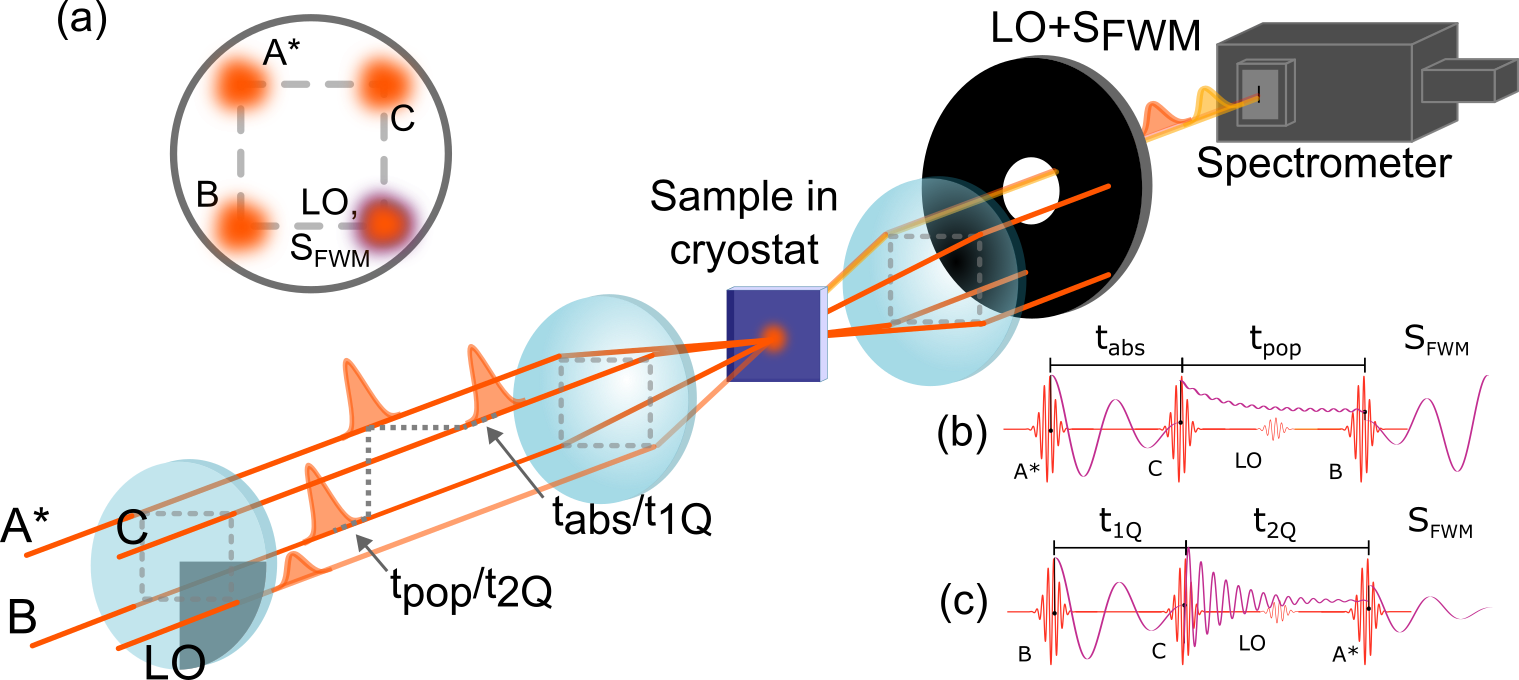}
    \caption{(a) The geometry of the excitation pulse-train beam pattern (red) and the resonant four-wave mixing signal (SFWM, yellow-orange), detected by interference with a local oscillator (LO). We use the BoxCARS beam geometry, in which three pulse trains (A, B, C) propagating along the corners of a square are focused onto the sample with a common lens, defining incident wave vectors $\mathbf{k_A}$, $\mathbf{k_B}$, and $\mathbf{k_C}$. The LO beam, on the fourth apex of the incident beam geometry, co-propagates with $S_{FWM}$ with the wave vector imposed by the chosen phase-matching conditions. (b) By changing the time-ordering of the pulse sequence, we measure two distinct nonlinear responses: (b) 1Q rephasing signal and (c) 2Q non-rephasing signal. 
}
    \label{fig:setup}
\end{figure}

The 2DES method resolves the nonlinear optical response of a system—coherent radiation generated by a nonlinear polarization induced by three phase-controlled femtosecond pulses—along two correlated energy axes: excitation and emission energies. Fig.~\ref{fig:setup}(a) shows the schematic representation of the experimental setup, in which $A^*$, $B$ and $C$ represent the excitation pulses, incident on the sample with their respective wavevectors -  $\mathbf{k_A}$, $\mathbf{k_B}$, and $\mathbf{k_C}$. The excitation energy response is extracted from a time-domain coherent excitation measurement, while the emission energy response is obtained via spectral interferometry with a fourth replica pulse, LO. This approach effectively reconstructs the spectral structure observed in the absorption spectrum, revealing diagonal peaks that represent optical-transition autocorrelations. Additionally, off-diagonal cross peaks emerge in the presence of correlations between different optical transitions. Importantly, excited-state absorption (ESA) contributions—such as those arising from multiexciton states—can be obscured in linear or incoherent nonlinear measurements like transient absorption but are distinctly identifiable in a 2D coherent spectral experiment. The multidimensional spectrometer used in this study is based on the Coherent Optical Laser Beam Recombination Technique (COLBERT)~\cite{turner2011invited}.

\begin{figure}
    \centering
    \includegraphics[width=8.5cm]{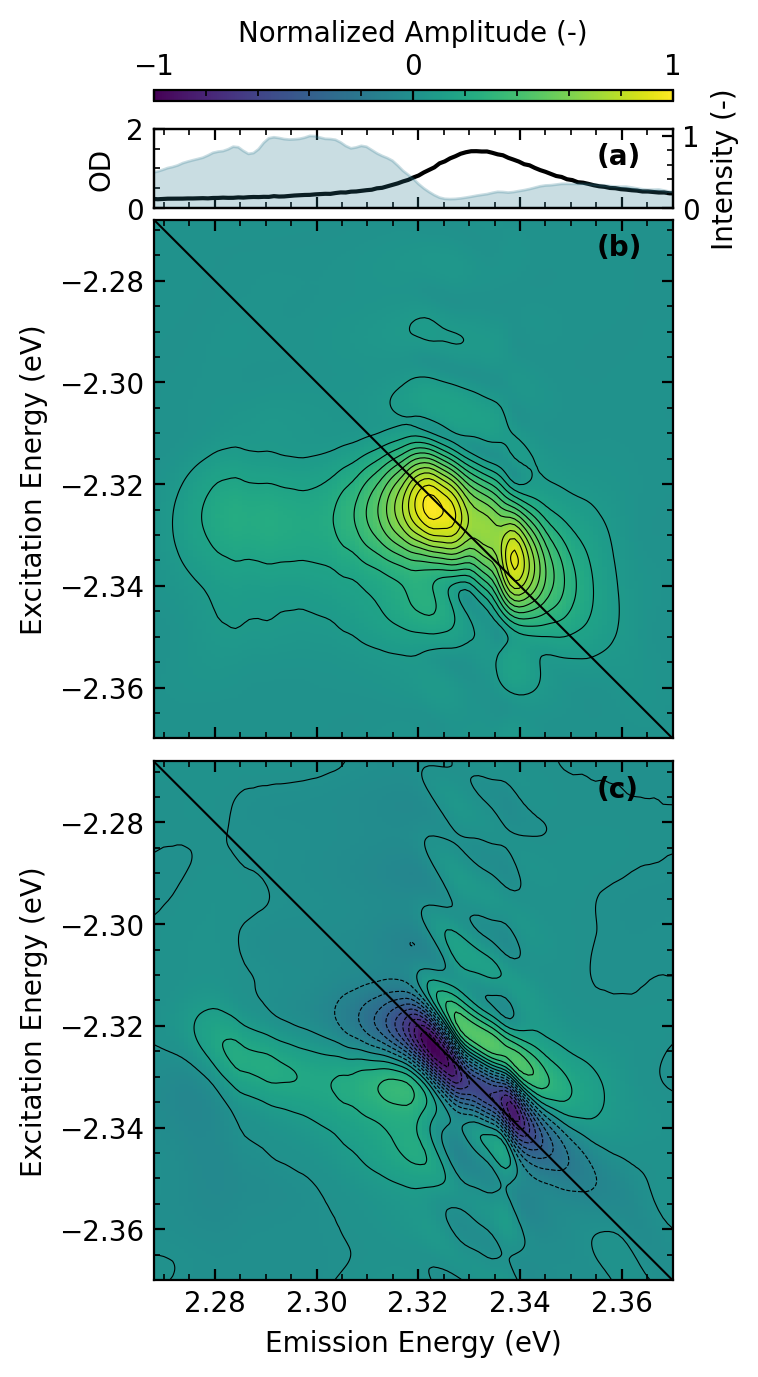}
    \caption{(a) The absorption spectra and pump laser spectra used during the experiments. (b) Absolute and (c) real one-quantum rephasing 2D coherent spectrum of \ce{(F-PEA)2PbI4} measured at 7K.}
    \label{fig:1Q}
\end{figure}

By modifying the sequence of incident pulses, as shown in Figs.~\ref{fig:setup}(b) and (c), our experimental setup allows us to conduct two distinct measurements: one-quantum (1Q) and two-quantum (2Q) correlations. The 1Q total-correlation measurement probes the transitions between the ground state and the first set of excited states, as well as subsequent excitations to higher states via an intermediate photo-excited population. In contrast, the 2Q non-rephasing measurement directly probes coherences between one-quantum and two-quantum states, without involving a population term. This makes it a more selective approach for detecting coherences associated with higher-lying excited states. Further details on these experimental schemes can be found elsewhere~\cite{turner2011invited} and in section B of the SI.

We first focus on the one-quantum experiments. We note that the chosen experimental scheme, where the emitted signal is acquired at $\Vec{k}_{sig}=-\Vec{k}_{a}+\Vec{k}_{b}+\Vec{k}_{c}$ (see Fig.~\ref{fig:setup}(b)), allows us to measure the 1Q rephasing spectrum, absolute and real parts of which are shown in Fig.~\ref{fig:1Q}(b) and (c) respectively. The pump laser spectrum which covers all the excitonic features in the absorption spectra of \ce{(F-PEA)2PbI4} is shown in Fig.~\ref{fig:1Q}(a). We observe two main features along the diagonal, corresponding to the two dominant resonances in the linear spectrum, which we label as $X_1$ and $X_2$. These resonances are obscured by the broad inhomogeneous broadening in the linear spectrum, while distinctly visible in the 2DES-1Q spectrum. The lineshapes of the features along the diagonal can be analyzed to estimate the exciton-phonon interactions and many-body scattering processes, which are discussed elsewhere~\cite{srimath2020stochastic}. Note that the absorption and PL data from Fig.~\ref{fig:abs_PL} were taken with a different spectrometer than the 2DES measurements. A slight energy discrepancy ($\approx$ 0.02 eV or 5nm) can be observed in the absorption from Fig.~\ref{fig:abs_PL}(a) and Fig.~\ref{fig:1Q}(a), which falls within the calibration resolutions of the two spectrometers. 

Now, we focus on the prominent off-diagonal feature, red-shifted from the diagonal. We attribute this to excited-state absorption (ESA) from the 1Q states to higher-lying 2Q transitions, consistent with previous studies~\cite{cundiff2009optical, dey2015biexciton, thouin2018stable}. This assignment is further supported by the observed relative $\pi$ phase shift in the single exciton emission within the real 1Q spectra (see Fig.~\ref{fig:1Q}(c)).

Moreover, this off-diagonal feature in the 1Q rephasing correlation spectrum provides a second approach for estimating the biexciton binding energy ($E_{B_1}$). Specifically, this corresponds to the energy difference between the excitation energy (ground state to exciton transition) and the emission energy of the ESA feature (exciton to biexciton transition). A critical factor in this estimation is precisely determining the ESA feature's peak position. To achieve this, we perform a gradient calculation on the absolute 1Q spectra, generating a vector map that allows for a more accurate localization of the ESA peak, see Fig. S3 and section C in the SI for more details. The resulting biexciton binding energy is estimated to be $E_{B_1} \approx 50$\,meV, approximately 7-10 meV higher than the estimation obtained from photoluminescence (PL) measurements. While we consider this a more precise estimate of $E_{B_1}$, we acknowledge that the feature lacks a well-defined peak. Consequently, substantial uncertainty remains in determining the peak position due to the extended tail of the feature toward the diagonal of the correlation map, even with gradient analysis.

\begin{figure*}
    \centering
    \includegraphics[width=17cm]{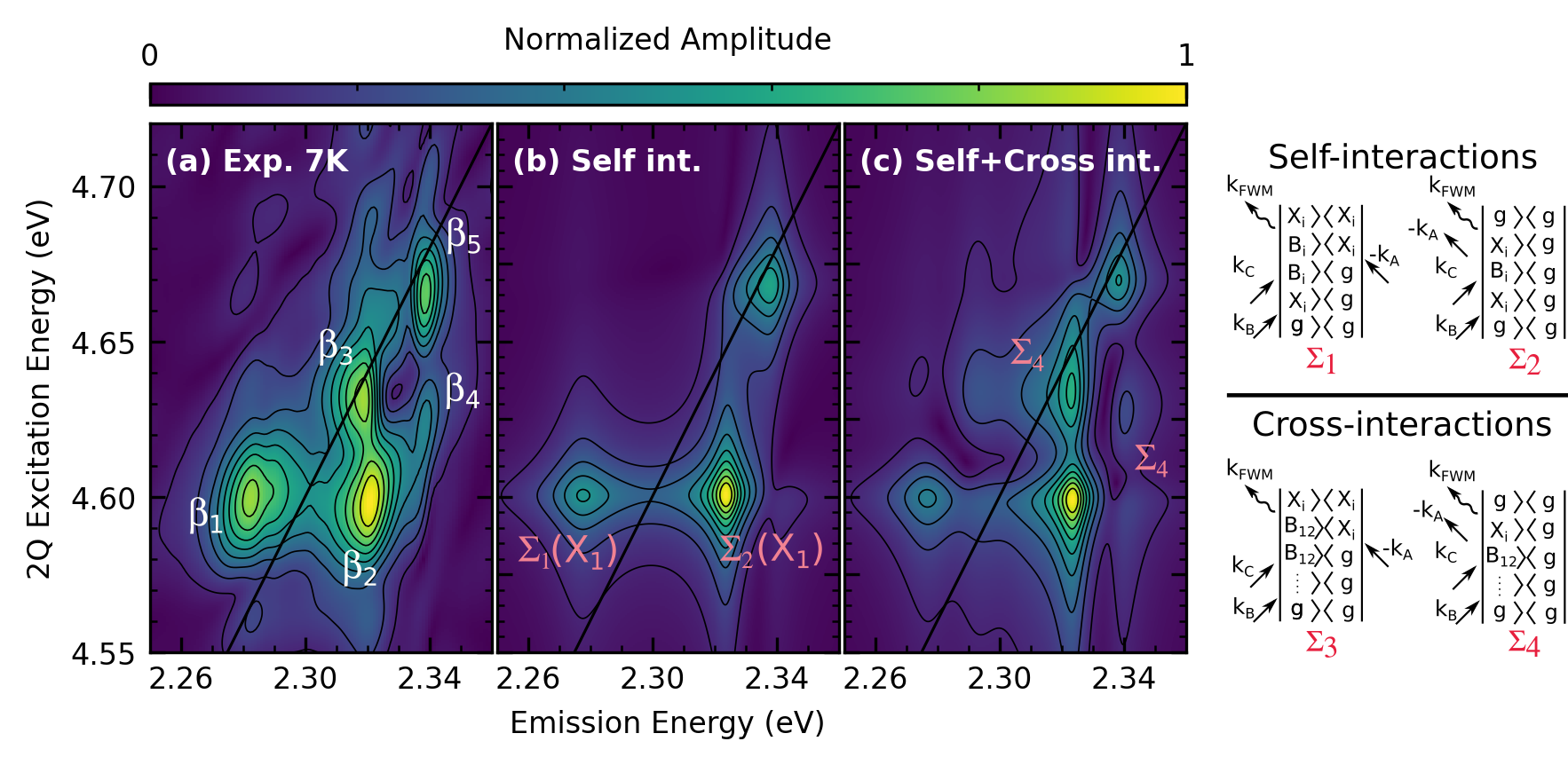}
    \caption{Two-quantum nonrephasing 2D coherent absolute spectrum of \ce{(F-PEA)2PbI4} measured at (a) 7K. Simulated two-quantum response considering two excitons ($X_1 =2.324$ eV and $X_2 =2.339$ eV) and the following Feynman pathways: (b) $\Sigma_{1,2}$ and (c) $\Sigma_{1,2,3,4}$. The biexciton binding energy for exciton 1  ($E_{B1} = 47.6$ eV) is determined from experiment. The binding energy for biexciton 2 ($E_{B2}$) and the mixed biexciton ($E_{B12}$) binding energy were set to, $E_{B2} = 10$\,meV and $E_{B12} = 33$\,meV. The Feynman pathways considered in the simulation are shown, where $X_{i=1,2}$ and $B_{i=1,2}$ represent the exciton and biexciton.}
    \label{fig:2Q}
\end{figure*}

Additionally, two distinct resonances along the diagonal indicate the presence of two separate excitonic species in this material. Each of these states is expected to have associated two-quantum states and possibly a \textit{mixed} two-quantum state. In fact, such states could contribute to the extended tail of the ESA feature and the broader linewidth in the PL spectrum. However, we lack direct access to the complex multi-particle correlations inherent in these systems due to the absence of clear spectroscopic signatures in both linear and one-quantum 2DES spectra. To further explore such correlations, we employ two-quantum 2DES methods.

In a two-quantum rephasing experiment, the emission signal is detected at $\Vec{k}_{sig} = \Vec{k}_{b}+\Vec{k}_{c}-\Vec{k}_{a}$, see Fig.~\ref{fig:setup}(c). This results in a rephasing spectrum that correlates the energy of two-quantum states with the energy of one-quantum excitations. The experimentally measured two-quantum (2Q) correlation spectrum (absolute value) of the sample at 7\,K is presented in Fig.~\ref{fig:2Q}(a). Five distinct and clearly visible features, labeled $\beta_1$ -- $\beta_5$, can be identified.

Before assigning these observed features, we first examine the expected signatures in a typical 2Q correlation map. If a two-quantum state exists with zero binding energy relative to a given one-quantum state, a feature should appear along the diagonal line $E_{excitation}=2E_{emission}$.  However, if there is a finite binding energy due to attractive interactions, the resonance is expected to appear below this diagonal. The shift from the diagonal quantifies the binding energy. This resonance can be understood through an excitation pathway in which the first two pulses generate coherence between the ground state and a two-quantum state, while the third pulse projects onto coherence between the one-quantum and ground states, emitting the measured four-wave mixing signal. This process is represented by the double-sided Feynman diagram $\Sigma_2$ in Fig.~\ref{fig:2Q}. Another possible excitation pathway, corresponding to the employed pulse sequence, is represented by diagram $\Sigma_1$, which leads to the emission at the energy of the exciton-biexciton coherence, and thus is expected to appear to the left of the diagonal. 

Returning now to the experimental map, we deduce that features features $\beta_1$ and $\beta_2$ correspond to diagrams $\Sigma_1$ and $\Sigma_2$ respectively, and are associated with the biexciton $B_1$ of the exciton $X_1$, the dominant peak in the linear absorption spectrum. Similarly, feature $\beta_5$, which corresponds to nonlinear signal emission at the energy of $X_2$, is assigned to diagram $\Sigma_2$ with biexciton $B_2$. 
To further understand the 2Q signatures, we simulate the 2Q coherent response by analyzing the Liouville-space pathways \cite{yang2007two}. 
We utilize the expressions for the emitted field, S$_{III}$ (2Q component), derived using time-dependent perturbation theory assuming that the pulses are in the impulsive limit. 
Based on the two dominant excitonic signatures observed in the 1Q spectra, we assume singly and doubly transitions for two distinct excitons ($X_1=2.324$\,eV and $X_2=2.339$\,eV). The model is similar to what has been described for the heavy and light holes in GaAs~\cite{yang2008isolating}. All the expressions are shown in section D of the SI.
For simplicity, we do not include complex multiexciton interactions~\cite{rojas2023many, srimath2020stochastic, stone2009two} (besides the biexciton binding energy) or inhomogeneous broadening, however, as shown below this minimal model is enough to describe the biexcitonic landscape.

To reproduce the observed features, we incorporate biexciton binding energies of approximately $E_{B_1}\approx47.6$\,meV and $E_{B_2}\approx10$\,meV. These values can also be estimated directly from the correlation map by measuring the positions of $\beta_2$ and $\beta_5$. The displacement of these features from the diagonal along the 2Q excitation axis provides an accurate estimate of the binding energy. The simulated 2Q correlation map considering all the self-interactions of $X_1$ and $X_2$, and associated with diagram $\Sigma_1$ and $\Sigma_2$ is shown in Fig.~\ref{fig:2Q}(b).

It can be seen from Fig.~\ref{fig:2Q}(b) that self-interactions of $X_1$ and $X_2$ alone are not sufficient to reproduce the experimental map. Specifically, we cannot rationalize features $\beta_3$ and $\beta_4$. To account for these features, we consider additional excitation pathways shown as diagrams $\Sigma_3$ and $\Sigma_4$, which contain excitation of coherence between the ground state and a \textit{mixed} two-quantum state $B_{12}$. This corresponds to a biexciton that forms due to the binding of $X_1$ and $X_2$ due cross-interactions. As it can be seen in Fig.~\ref{fig:2Q}(c) pathways $\Sigma_3$ and $\Sigma_4$ accurately reproduce the $\beta_3$ and $\beta_4$, confirming the presence of a biexcitonic state with mixed character. To reproduce the energetic location of the $\beta_3$ in the 2Q map, we evaluate the mixed biexciton binding energy, $E_{B_{12}}$, to be approximately 33\,meV, see Fig. S4(a) and section C of the SI for more details. 

\section{Discussion}
\begin{figure}
    \centering
    \includegraphics[width=1\linewidth]{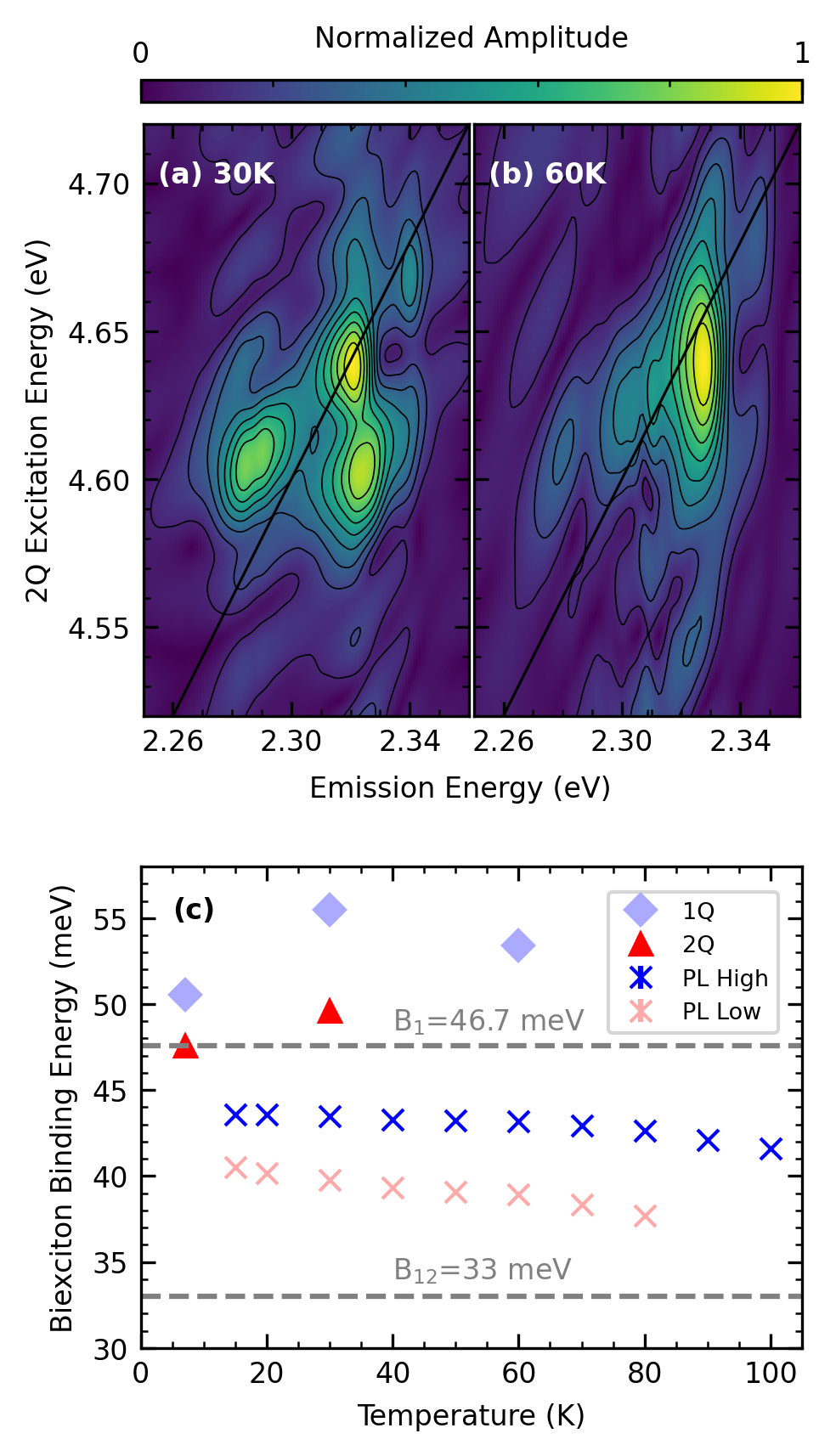}
    \caption{Two-quantum nonrephasing 2D coherent absolute spectrum of \ce{(F-PEA)2PbI4} measured at (a) 30K and (b) 60K. (c) Biexciton binding energy calculated from, photoluminescence and one-quantum rephasing and two-quantum nonrephasing 2DES measurements.}
    \label{fig:biX_bind}
\end{figure}

Returning to the initial motivation behind performing two-dimensional electronic spectroscopy (2DES) measurements—obtaining precise and unambiguous estimates of biexciton binding energy—we emphasize that the two-quantum (2Q) measurements provide the clearest spectroscopic signatures associated with biexcitons among the three methods explored in this study. The 2Q correlation map reveals well-defined and distinct spectral features corresponding to various biexcitons, which arise due to self-interactions and cross-interactions between different excitonic states. This level of resolution and specificity is not as easily achieved with the other techniques.

By consolidating the biexciton binding energy estimates obtained from the three different methods, we find that $E_{B_1}$ is determined to be approximately 40\,meV from photoluminescence (PL) spectroscopy, 46.7\,meV from the 2Q spectrum, and 50\,meV from the one-quantum (1Q) spectrum. Although all of these values are relatively close, the estimate derived from the 2Q spectrum is considered the most reliable. This is because 2Q spectroscopy minimizes ambiguities related to spectral broadening and overlapping resonances, providing a more direct measurement of biexciton binding energies. 

The temperature dependence of biexciton binding energy, as determined from the three different methods, is shown in Fig.~\ref{fig:biX_bind}(c), revealing an intriguing trend. Notably, the binding energy extracted from PL spectra is consistently lower than the values obtained from the other two methods across all measured temperatures. More significantly, biexciton binding energy exhibits a clear decreasing trend with increasing temperature. We focus first on the estimates obtained with lower excitation density, shown as \textit{PL Low} in the plot. Note that for this dataset, it was not possible to estimate binding energy beyond 80\,K due to lack of clear biexciton peak in the spectrum. While this reduction could be attributed to thermal fluctuations weakening the interaction strength between excitons, it is important to interpret these results with caution. The estimation of binding energy from PL spectra is particularly susceptible to uncertainties, primarily due to self-absorption effects and broad spectral linewidths, which become more pronounced at higher temperatures. These factors introduce challenges in accurately determining biexciton binding energy, underscoring the limitations of PL-based measurements in this context.

Figs.~\ref{fig:biX_bind}(a) and (b) present the 2Q spectra recorded at 30\,K and 60\,K, respectively. The 30\,K spectrum closely resembles that at 7\,K (Fig.~\ref{fig:2Q}(a)), with a nearly identical estimated binding energy. At 60\,K, however, the spectrum undergoes a significant transformation, with feature $\beta_3$ becoming more prominent. This makes it difficult to estimate the binding energy of $B_1$ and, consequently, to determine its temperature dependence. Nevertheless, features $\beta_1$ and $\beta_2$ remain at similar positions, suggesting minimal changes in the biexciton binding energy—contrary to what PL measurements imply. This indicates that enhanced disorder has little effect on the exciton binding energy at temperatures below 60\,K. However, such effects may become more pronounced above 100\,K, a regime beyond the scope of this manuscript.

Examining the binding energy obtained via the PL method at higher excitation densities, we find consistently higher values compared to those at lower densities. This behavior may result from many-body effects or state filling, which alter excitonic interactions. However, due to the inherent limitations of the PL-based method, this observation cannot be directly verified. Similarly, the biexciton binding energies extracted from the 1Q spectrum are also higher than those from the 2Q spectrum. This difference is noteworthy, as the one-quantum nonlinear response involves a population term, whereas the 2Q response arises purely from coherences (see diagrams in Fig.~\ref{fig:2Q}). Thus, the 2DES data supports the PL-based suggestion that higher excitation densities may enhance biexciton binding.

From a photophysical perspective, however, binding energy is expected to decrease due to screening effects. These effects also manifest as excitation-induced shifts and characteristic dispersive lineshapes in the 1Q spectrum, as discussed in our previous work~\cite{srimath2020exciton}. The impact on biexciton binding warrants further investigation, particularly through studies of biexciton population dynamics, which require a $\chi^{(5)}$ process induced by a five-pulse interaction.

The trends discussed above regarding intensity and temperature, summarized in Fig.~\ref{fig:biX_bind}(c), raise several pertinent questions. While we have only a few indicative answers, these findings inspire more focused research to unravel the complex photophysics of biexcitons. That said, our study highlights the utility of 2Q correlation maps in both improving the accuracy of binding energy estimates and identifying signatures of self- and cross-interactions between excitons.

One may, however, be intrigued by the absence of signatures of this mixed state in the 1Q and PL spectra. Notably, the ESA feature exhibits an extended tail, hinting at the possible presence of another feature, as has been observed for \ce{(PEA)2SnI4}.\cite{rojas2023many} Moreover, as discussed in Section II, the biexciton feature in the PL spectrum has a substantially broader linewidth compared to the exciton. This broadening could result from overlapping emissions of two co-existing biexcitons, $B_1$ and $B_{12}$. The binding energy of $B_1$ is small enough to strongly overlap with the emission from $X_1$. In fact, the binding energy estimated from PL falls between the values of $B_1$ and $B_{12}$ (indicated by dashed lines in the figure), suggesting that the PL spectrum averages over both states. 

Additionally, the 2Q spectra in Fig.~\ref{fig:biX_bind}(b) reveal that the relative intensity of the feature associated with $B_{12}$ increases with increasing temperature. Consequently, as the PL binding energy shifts towards the value of $B_{12}$ at elevated temperatures, this suggests an increased contribution from the mixed state. However, the underlying mechanism driving this trend remains unclear, and its connection to PL variations is speculative. Nevertheless, our results provide definitive evidence of the mixed state's presence in 2Q spectra.

The identification of multiple biexcitonic states with distinct characters and binding energies has significant implications for the photophysics of Ruddlesden-Popper metal halides and, more broadly, for metal halide perovskites. These technologically relevant materials are well known for their strong electron-lattice coupling effects, which impart polaronic character to their photo-excitations. Previously, we discussed how the peculiar optical lineshape at the exciton energy—such as the one shown in Fig.~\ref{fig:1Q}(a)—arises from these unique lattice interactions. The distinct resonances, labeled $X_1$ and $X_2$ for convenience, likely represent exciton polarons with different lattice coupling strengths~\cite{srimath2020exciton}. However, an alternative interpretation in the literature suggests that these multiple resonances are simply phonon replicas of a single exciton state~\cite{straus2018electrons}. In this context, our observation of the mixed biexciton state supports our perspective that $X_1$ and $X_2$ are indeed two distinct states, capable of Coulombic cross-coupling to form mixed biexcitons. furthermore, $X_1$ and $X_2$ exhibit distinct self-interactions, as reflected in the significantly different binding energies of $B_1$ and $B_2$. Since these excitons are uniquely dressed by lattice phonons with varying degrees of anharmonicity~\cite{thouin2019phonon, srimath2020exciton}, our findings suggest that lattice dressing plays a crucial role in shaping multi-excitonic correlations. 

Finally, we note that a similar study on another Ruddlesden-Popper metal halide, \ce{(PEA)2PbI4}, revealed evidence of both attractive and repulsive self-interactions but no dominant signatures of cross-coupling~\cite{thouin2018stable}. In contrast, the material variant examined here, featuring a substituted organic cation, exhibits slightly different biexcitonic couplings. Recent findings suggest that variations in the organic cation induce subtle modifications in the lattice degrees of freedom. A more systematic and detailed investigation is needed to understand how such lattice modulations influence the many-body interactions of excitons.


\section{Conclusions \label{sec:concl}}



We investigated the spectroscopic signatures of biexcitons in the two-dimensional hybrid semiconductor \ce{(F-PEA)2PbI4} using three different methods. The widely used PL-based technique, which identifies biexcitons through red-shifted features in the PL spectrum at elevated densities, is hindered by reabsorption and spectral broadening, making its biexciton binding energy estimates unreliable. Coherent nonlinear spectroscopy provides a more robust alternative. In the 1Q rephasing spectrum, we identify excited-state absorption features as biexciton signatures. However, significant spectral overlap still prevents precise binding energy estimation. The 2Q spectrum, which maps correlations between one-quantum and two-quantum excitation energies, emerges as the most accurate method for determining the biexciton binding energy. Furthermore, it reveals the diversity of biexciton states arising from multiple excitonic resonances. Our findings show that biexciton binding energy varies depending on the self-interaction of distinct excitonic states. Additionally, we identify a clear spectral feature supporting the existence of a mixed biexciton state resulting from exciton cross-coupling. Lastly, by comparing our results with existing literature on other Ruddlesden-Popper metal halide variants, we observe that biexcitonic properties may be influenced by organic cation substitution in the material.


\begin{acknowledgments}
ARSK acknowledges funding from the National Science Foundation CAREER grant (CHE-2338663), start-up funds from Wake Forest University, funding from the Center for Functional Materials at Wake Forest University. Any opinions, findings, and conclusions or recommendations expressed in this material are those of the authors(s) and do not necessarily reflect the views of the National Science Foundation.   
The experimental data collection, analysis, and the writing of corresponding manuscript sections by ERG were supported by the National Science Foundation (DMR-2019444).  CSA acknowledges funding
from the Government of Canada (Canada Excellence Research Chair CERC-2022-00055) and the
Institut Courtois, Facult\'e des arts et des sciences, Universit\'e de Montr\'eal (Chaire de Recherche
de l’Institut Courtois).

\end{acknowledgments}

\section*{Author Contributions}
The measurements were performed by KAK and ERG under the supervision of CSA and ARSK. The samples were prepared by MGD under the supervision of JPCB, and contributed to the analysis of the linear optical spectra. The 2DES data was analysed by KAK and ERG under the supervision of ARSK. ARSK wrote the original draft with contributions from KAK and ERG, and all the authors contributed to editing the manuscript. The project was conceived and co-ordinated by ARSK. 






%

%

\end{document}